# Physical features of small disperse coal dust fraction transportation and structurization processes in iodine air filters of absorption type in ventilation systems at nuclear power plants

O. P. Ledenyov, I. M. Neklyudov, P. Ya. Poltinin, L. I. Fedorova

*National Scientific Centre Kharkov Institute of Physics and Technology, Academicheskaya 1, Kharkov 61108, Ukraine.*

The research on the physical features of transportation and structurization processes by the air-dust aerosol in the granular filtering medium with the cylindrical coal adsorbent granules in an air filter of the adsorption type in the heating ventilation and cooling (*HVAC*) system at the nuclear power plant is completed. The physical origins of the coal dust masses distribution along the absorber with the granular filtering medium with the cylindrical coal granules during the air-dust aerosol intake process in the near the surface layer of absorber are researched. The quantitative technical characteristics of air filtering elements, which have to be considered during the optimization of air filters designs for the application in the ventilation systems at the nuclear power plants, are obtained.

PACS numbers: 45.70.-n, 45.70.Mg, 87.16.dp .

Keywords: air filtering, iodine air filter (*IAF*), air – dust aerosol, granular filtering medium (*GFM*), cylindrical coal adsorbent granules, high disperse coal dust fractions, coal dust precipitations, chemical absorber, heating ventilation and cooling (*HVAC*) systems, radioactive chemical elements, nuclear power plant (*NPP*).

## Introduction.

The air filtering with the application of iodine air filters (*IAF*) with the coal adsorbent granules allows to prevent the possible radioactive contamination of environment by establishing the technological process of air cleaning from the radioactive chemical elements, produced at the nuclear power plant (*NPP*). During the many years operation of the *IAF* in the heating ventilation and cooling (*HVAC*) systems at the *NPP*, there is a systematic non-monotonous increase of the magnitude of aerodynamic resistance by the *IAF*, which results in a substantial decrease of the *IAF*'s operational effectiveness on the order of magnitude and higher. The main physical problems, causing the described phenomena, were described in [1, 2]. It was found that in the thin filtering layer, situated in near the surface region of the granular filtering medium (*GFM*), where the air stream flow experiences the change from the regime of the free flow to the regime of the structured channels flow, the coal granules are partially destroyed. As a result, the high disperse coal fraction is created, because of the dynamic fluctuations of pressure, accompanied by the micro- and macro- transpositions of structural elements in the granulated filtering medium in an air filter. The high disperse coal fraction participates in the two interconnected dynamic processes:
1) The high disperse coal dust fraction transportation along the granular filtering medium in an air filter;
2) The high disperse coal dust precipitations structurization in the granular filtering medium in an air filter, creating some barriers on the way of aerosol flow through the granular filtering medium in an air filter, because of a decrease of the effective operational cross-section of adsorber in an air filter. The overall impact by the above listed two physical processes appears in the certain limitation of *IAF*'s technical characteristics.

In the present research, based on the modeling experiment [2], the detailed analysis of obtained experimental results is conducted. The physical origins of observed physical processes in the granular filtering mediums with the high disperse coal dust fractions are clarified. Let us note that, despite a big number of completed research programs over the recent years, the high disperse coal dust mediums represent a subject of considerable research interest, and are researched intensively [3, 4]. The given research is quite complex, because of both:
1) The complexity of researched physical processes in the granular filtering mediums with the high disperse dust fractions, which can not be characterized by the simple model representations; and
2) The necessity of consideration of nano-structured systems in view of a growing number of their technical applications in the fields of nuclear physics and engineering, chemical physics, and pharmaceutical chemistry.

## Analysis of experimental results. Design of air filter with granular filtering medium with cylindrical coal absorbent granules.

In [2], the detailed research results on the dynamics of distribution of the high disperse coal dust fraction along the granular filtering medium with the cylindrical coal absorbent granules during the process of the air-dust aerosol flow in the near the surface layer of absorber's input in an air filter, is completed. The *IAF* includes the granular filtering medium, consisting of the absorber with the cylindrical coal absorbent granules with the diameter $d \approx 1,8\ mm$ and length of $h \approx 3,2\ mm$. The absorber's structure has no any particular geometric order. The distribution and orientation of granules is randomly established during the process of absorber filling. The air-dust aerosol flows over the conditional air channels between the granules, which are also randomly distributed. The average density of granular filtering medium in relation to the weighted density of granular material is around $\approx 77\%$. The granular filtering medium consists of the cylindrical coal absorbent granules, which create the porous structure with the complex configuration of empty cavities and constrictions with the small characteristic dimensions in comparison with the conditional dimensions of macroscopic air channels in an air filter (in more than $10^3$ times smaller). The penetration of air-dust aerosol with the radio-nuclides from the channels to the filling filtering medium with the porous structure with the empty cavities and constrictions takes place due to the diffusion process, and it is not connected with the aerosol's velocity in the air channels of an air filter. The aerosol's penetration ability into the cylindrical coal granules is much smaller than the aerosol's penetration ability into an air filter (the aerosol's penetration ability into an air filter has an opposite value to the aerodynamic resistance). Therefore, the aerosol's transportation mainly takes place in the air channels between the cylindrical coal granules in the granular filtering medium, but it doesn't occur due to the infiltration through the bodies of cylindrical coal granules. The parallel ordering of cylindrical coal granules may result in a dense packaging of cylindrical coal granules, creating a rhombic elementary cell in its cross-section, including the one cylindrical coal granule and two empty spaces. The porosity of granular filtering medium can be expressed as the relation of the free space volume to the absorber's volume, and it is close to $\Omega \approx 18,5\%$. There may be a significantly bigger number of possible configurations of the granules ordering into the structure with the one cylindrical coal granule and one empty space per a quadratic elementary cell. The porosity of such a structure is slightly increased up to $\Omega \approx 21,5\%$. In the both cases, the air blow towards the directions of transverse axes of filtering structure is not possible, and the aerosol transportation is going along the longitudinal axis. In the general case, the quadratic structure with the parallel orientation of elements is unstable in relation to the transition to the more dense rhombic structure. However, in the quadratic structure, it is possible to achieve the rotation of upper layer in such a way that it becomes parallel to the horizontal substratum, but its elements have the changed orientation of longitudinal axis. At the rotation on the angle $\varphi \geq \arcsin\left(\dfrac{d}{h}\right) \approx 54°$, this structure becomes stable, and it can't transit to the rhombic structure spontaneously. In the considered case, a number of possible configurations with various rotation angles is significantly bigger, comparing to the case of a simple quadratic cell. The most stable structure appears at the average rotation on the angle of $\pi/2$. The rotations of layers don't result in a change of the porosity of structure, leaving it equal to the value of porosity of a simple quadratic cell. However, there is a probability that the aerosol stream can flow along the one additional axes from all the axes of the structure. In this structure, there may be the rotation of elements of layers in the vertical plane, resulting in the absolute penetrability of structure towards all the three axes. It looks like that the similar structure is realized in the practical cases as shown in Fig.1. There may be some additional empty cavities in the transition places from the one allowable element's orientation to the another one, resulting in an increase of absorber's porosity up to $\Omega \approx 1,5\%$, that is up to the experimentally observable value, which is equal to $\Omega \approx 23\%$ approximately.

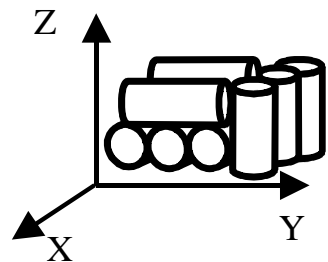

*Fig. 1. Assumed local geometric structure of granular filtering medium with cylindrical coal granules in air filter.*

The total thickness of absorber layer is $L=30\ cm$, and all the orientations of elements-granules are possible, thus, the most probable conditional aerosol transportation length is equal to the length of rout along the ribs of subsequently distributed structural cubes with the common diagonal $L$ in Fig. 2. The average length of rib of an elementary structural cube may be considered as equal to the geometric mean granule dimension $(d \cdot h)^{1/2}$, and its volume is $V_0=(d \cdot h)^{3/2}$. The conditional number of structural layers is $n = L/(d \cdot h)^{1/2}$, and conditional channel length to transfer the air-dust aerosol is $L^* = 3^{1/2} \cdot (d \cdot h)^{1/2} \cdot n = (3)^{1/2} \cdot L$. The average number of channels is approximately equal to $N=S/(d \cdot h)$, where $S$ is the square of cross-section of absorber. The average cross-section of conditional cylindrical channel can be written as $S_C = \pi \cdot R^2$, where $R$ is the average channel's radius. The radius can be found, going from the equality between the free volume of porous structure and the full volume of air channels. Considering that the volume of single air channel is $S_C \cdot L^*$, hence the full volume of all the air channels is $V_C = N\ S_C L^* = \Omega\ V_F$, where $V_F$ is the volume of air filter. Therefore, the radius is $R = \sqrt{\dfrac{\Omega\ dh}{\pi\ \sqrt{3}}}$. This expression gives an opportunity to estimate the cross-cut dimension of air channel, taking to the consideration the granule's dimension and the porosity of granular filtering medium layer within the absorber.

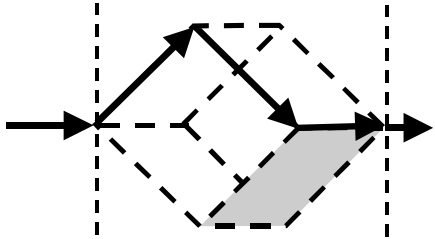

*Fig. 2. Conditional air-dust aerosol flow by edges of cubic structure in granular filtering medium in air filter.*

## Physical features of air flow in granular filtering medium with cylindrical coal absorbent granules in air filter.

The character of aerosol stream flow dynamics in an air filter is defined by the value of the *Reynolds* number $Re \approx 2 \cdot R \cdot \upsilon\ /\ \nu$. The calculation shows that, at the described conditions of an experiment, up to the maximal air stream magnitude, the *Reynolds* number $Re$ doesn't exceed the characteristic critical threshold of turbulent air stream flow ($Re_c = 2300$), hence the air – dust aerosol flow must be conducted by the laminar flow ($\upsilon$ is the aerosol stream velocity, $\nu$ is the kinematic coefficient of aerosol viscosity, $\nu = \mu/\rho_{air}$, where $\mu$ is the dynamic coefficient of viscosity, $\rho_{air}$ is the density of aerosol). Let us note that the aerosol flow velocities distribution across the cross-cut of an air channel is not homogenous. At the laminar flow, the aerosol flow velocity distribution obeys the well-known parabolic law, when the velocity is equal to the nil on the surface of the channel's walls (granules), but the velocity reaches its maximal value $\upsilon_m$ at the center of channel. The average velocity of aerosol stream flow is $\upsilon = \upsilon_m/2$. At the turbulent aerosol flow, the distribution of the longitudinal velocity across the cross-section of air channel is relatively simple. At the close proximity to the channel's wall, the velocity reaches its maximal value and remains constant at the cross-section of channel. In the case of laminar flow, the aerodynamic resistance by the air channel is [5, 6]

$$\Delta P = \dfrac{8 J \mu L^*}{\pi R^4}$$

where $J = \Omega\ S\ V_F$ is the volumetric outgo of aerosol per unit of time. As it follows from this expression, there is a linear dependence between the pressure difference and the aerosol stream flow. However, the universal dependence $\Delta P \propto J^\kappa$ ($\kappa \approx 1,5$) is true for all the values of aerosol stream flow, starting from the very small levels and up to the maximal reachable levels of dust content, as shown in Fig. 4 in [2]. In an agreement with the fluid dynamics theory, the value $\kappa > 1$ is characteristic for the transition region from the laminar flow to the turbulent flow and the conditions of turbulent filtration [7]. As it is known, the developed turbulent aerosol stream flow is well characterized by the value $\kappa = 2$, hence the degree of radius $R$ is decreased to the value of $\approx 2,5$. In [2], it appears that the design features of air filter's structure make a main contribution to an increase of aerodynamic resistance's magnitude, and its deviation from the magnitude of aerodynamic resistance at the laminar flow, resulting in an origination of so-called local additional aerodynamic resistances. It is necessary to make a certain work to change both the direction of aerosol flow as well as the direction of aerosol velocities during the aerosol flow over the constrictions in an air channel (the convergent flow), the enlargements in an air channel (the divergent flow), and the turns of air channel in an air filter. This process is naturally connected with the energy dissipation process in the viscous medium. In view of the air filter's physical properties, it is necessary to consider the variables connected with both the dimensions, characterizing the aerosol stream flow in an averaged air channel in the granular filtering medium (micro-dimension) as well as the dimensions, characterizing the aerosol flow around some macro-structures such as the sides of cube (macro-dimension) with the purpose to calculate the *Reynolds* number in Fig. 2. This results in an increase of the *Reynolds* number. In these cases, it is possible to connect the small value of *Reynolds* number $Re$, obtained at the calculation on the aerosol stream flow through the conditional air channel, with both the presence of additional aerodynamic resistances and the change of type of dependence $\Delta P(J)$ at the transition from the laminar flow to the turbulent flow by the aerosol stream.

## Physical features of air – dust aerosol flow in granular filtering medium with cylindrical coal absorbent granules in air filter.

In the analyzed case, the air-dust aerosol flows through an air filter. As it was shown in [2], the degree of filling of empty cavities by the coal dust precipitations at the sub-surface layer of absorber in relation to the available free volume can be up to *100%*. During this process, some fraction of coal dust precipitates and structures, the rest flows along the absorber.

In [2], the modeling experiment included the renewable source of coal dust, situated in a container with the thickness of *2cm*. This container was filled with the cylindrical coal absorbent granules as well as the small disperse coal dust fractions, and it was placed in front of an absorber. The small disperse coal dust fractions were created by the destruction of coal dust granules down to the dimensions below of *10 μm*, hence they were identical to the small disperse coal dust fractions, appearing at the real operation of an absorber in an air filter. Concerning the fractional content of small disperse coal dust, it is possible to make a general clarification that the granulation of monolithic volumetric material results in the logarithmically normal distribution of particles dimensions [8]. In distinction, the crushing of cylindrical coal granules in the absorber of the type of *CKT-3* during the process of small disperse coal dust fractions creation results in an appearance of a number of discrete dimensions of particles, because of the physical features of absorbent material. Therefore, the created small disperse coal dust fractions has the non-homogenous content in terms of particles dimensions. Let us emphasis that the fractional content of small disperse dust particles was not researched separately, however the distribution of dust masses along the length of absorber at the completion of modeling evidently confirms the described granulation character of small disperse dust particles during the crushing of cylindrical coal granules of the type of *CKT-3*.

The air - dust aerosol flow is a complex physical process, which is defined by the interaction between the various forces: the gravitation force, depending on the dust particles weight $F_g$; the capturing forces, depending on both the dynamic pressure force $F_P$ and the viscous friction force $F_S$; the adhesion forces $F_A$, interacting between the dust particles, and between the dust particles and the cylindrical coal granules; the centrifugal forces $F_I$, which define the dynamics of dust flow in the turn points of trajectory. The dynamic pressure force $F_P = \Delta P \cdot S$ has a main influence on the absorber as a whole, and it has a relatively small influence on the separate dust particles, hence it can be neglected. The evaluation shows that, at the big magnitudes of aerodynamic resistance $\Delta P$, the dynamic pressure force is comparable with the weight cylindrical coal granules in an air filter. The capturing forces are mainly defined by the magnitude of *Stocks* force $F_S = 3 \cdot \pi \cdot \mu \cdot \delta \cdot \upsilon_d$, where $\mu$ is the dynamic coefficient of viscosity, $\delta$ is the diameter of a particle, $\upsilon_d = \upsilon_r - \upsilon$ is the difference between the aerosol flow velocity and the particle's velocity. The small disperse coal dust transportation by the aerosol in an air channel is possible, when the magnitude of capturing force is bigger than the magnitude of gravity force, $F_S > F_g$. In the general case, there are complex dependences of considered parameters, going from the changing orientations of all the described physical forces. To avoid the unnecessary complications during the consideration of research problem, let us assume that the described physical forces are parallel. This situation is quite possible, because the aerosol stream flow is directed downwards, and it's direction coincides with the direction of gravity force. As it can be seen, the conditions of coal dust particle's tearing and capturing at the air channel's wall at the laminar aerosol flow are very specific, because the aerosol's velocity is small, and its value is close to the nil near the air channel's walls. Therefore, the big enough average aerosol velocities $\upsilon$ in an air channel are required with the purpose to capture the small disperse coal dust particles. There are the conditions at which the dust can be accumulated at the air channel's walls. The process of dust accumulation takes place inside the air channels with the alternate cross-sections such as the considered air channels formed in the granular filtering medium in an air filter, when the aerosol velocity decreases in the areas of channel's cross-section enlargement. The physical forces, connected with the difference of pressures between the aerosol's pressure near the wall of air channel and the aerosol's pressure in the center of air channel, have to be taken to the consideration in the case of dust particles with the relatively big surface area. These forces can tear a dust particle away from the air channel's wall and include it into the movement in the air-dust aerosol flow in an air filter. At the laminar regime at the initial stage, the small disperse coal dust fraction transports with the acceleration along the rectilinear part of air channel with the length of around $(d \cdot h)^{1/2}$ under the capturing force action, which has a parallel orientation in relation to the air channel. As we emphasized before, the acceleration magnitude depends on the difference between the aerosol's velocity and the dust particle's velocity, hence the acceleration magnitude decreases as the particle's velocity increases. The velocity of dust particle and the acceleration of dust particle are interconnected as expressed in the integral-differential equations

$$\vec{\upsilon}(\tau) = \vec{\upsilon}_0 + A \int_0^\tau \vec{\alpha}(t) dt , \qquad (1)$$

$$\vec{\alpha}(t) = \frac{\vec{F}_S}{m} = \frac{18\mu(\vec{\upsilon}_r - \vec{\upsilon}(\tau))}{4\delta^2 \rho} , \qquad (2)$$

where $\upsilon(\tau)$ is the vector of velocity by a dust particle at the time moment $\tau$, $A$ is the constant, $\alpha(t)$ is the vector of acceleration of a dust particle, depending on the time, $m$ is the mass of a dust particle, $\rho$ is the density of a dust particle, $\upsilon_r$ is the vector of aerosol velocity at the point $r$ in an air channel. To simplify this expression, it is possible to introduce the phenomenological mean free part by a dust particle of the type of $\ell_i$, which will be defined by the radius and length of air channel as well as by the dimension and material's density of dust particle. Then, the time moment $\tau_i$ will be equal to the time of existence of a dust particle of the type of $i$ in this particular state, and the force $F_S$ can be presented as a variable, which depends on the aerosol's velocity in an air channel. As it can be seen, the acceleration magnitude increases at the decrease of particle's diameter $\propto 1/\delta^2$, hence the big coal dust particles require much longer mean free path in comparison with the mean free path by the small coal dust particles, when reaching the same velocity. Let us emphasis that, in this approach, all the coal dust particles will finally reach the same velocity of transportation at the infinite mean free path in a relatively long air channel, which will be equal to the aerosol's velocity.

The interesting conclusion can be derived, taking to the consideration the fact that the coal dust particles movement can't be characterized as the movement without the collisions (this case has place, when the relative concentration of dust in the air is bigger than the one percent > *1%*). In this case, the transfer of energy and impulse from the one fraction of particles to another fraction of particles, differing by the particles dimensions, velocities and mean free paths, can be realized. Let us consider a well-known approach about the equal distribution of energies on the degrees of freedom of particles with the different energies and masses, which makes sense in the case of consideration of mixture of gases with the different molecular masses. This particles gas has the general characteristic temperature $T^*$, which is connected with the velocity of dust particles transportation, appearing at the action of aerosol stream flow. The dust particles experience the three dimensional movement, and get some effective temperature due to the turbulent aerosol stream flow, which is characterized by the multiple collisions of the dust particles with the channel's walls. In this case, the dust particles can be characterized by the *Maxwell* distribution and their effective temperature is

$$T^* \sim \frac{m_i \cdot v_i^2}{3 k_B} \quad (3)$$

where $v_i$ is the average velocity of particles of the type of $i$, $k_B$ is the *Boltzmann* constant. Then, the velocity of dust particles is equal to

$$v_i \sim \left( \frac{k_B \cdot T^*}{\pi \, \delta^3 \rho} \right)^{1/2} \quad (4)$$

As it appears, in this case, the average velocity of dust particles depends on the dimensions of dust particles. Therefore, the filtered dust particles are separated in the groups, depending on their dimensions and velocities, during the process of dust particles transportation in a researched air filter. This process is observed in the conditions of real experiment.

The centrifugal force defines the dynamics of particles movement in the region of turn points in an channel

$$F_i = \frac{\left(\frac{4}{3}\right) \pi \, \delta^3 \rho \, v^2}{R}$$

where $\rho$ is the density of dust particles material, $v$ is the dust particles velocity, $R$ is the air channel's radius, which is $R = d/2$ at the initial stage. The bigger dust particles with $F_I > F_S$, have collisions with the air channel's walls, resulting in an appearance of the transverse velocity in relation to the air channel's axis, and an origination of a series of subsequent collisions. Considering the fact that the air channel's walls with the cylindrical surfaces have the negative curvature in relation to the center of an air channel, hence there may be an effect, which will originate the "complication" of dust particles trajectories and their chaotic movement. This process doesn't depend on the presence of turbulent aerosol movement in an air filter.

Let us note that, in a general case, the presence of dust in the air - dust aerosol, must result in an increase of the dynamic viscosity of the air – dust aerosol in an agreement with the *Einstein* formula in eq. (5) [6]

$$\mu_{mixture} = \mu \left( 1 + \left(\frac{5}{2}\right) \beta \right) \quad (5)$$

where $\beta$ is the volumetric share of small disperse dust in the air - dust aerosol.

Moreover, it is necessary to take to the consideration the fact that there is a significant change of the average density of air - dust aerosol in comparison with the air density, if a few volumetric percentages of dust are present at least. Such air – dust aerosol creates the medium, which can be described by the equations of electro-dynamics, resulting in an increase of the *Reynolds* number proportional to its density $Re \propto \rho_{mixture}$. This impacts the character of air - dust masses movement, and results in an appearance of certain physical features at air – dust aerosol flow in an air filter.

## Physical features of dust structurization in granular filtering medium with cylindrical coal absorbent granules in air filter.

Let us emphasis that the negative curvature of surface appears in the both cases at the transversal movement by a dust particle as well as the longitudinal movement by a dust particle in the considered geometry of cylindrical coal granules positioning in an air channel in the granular filtering medium in an air filter. The loss of impulse by the dust particles with the subsequent effective self-structurization by the small disperse dust particles have place in the regions with the negative curvature of surface due to the complication of their trajectories of movement. This property is characteristic in relation to all the disperse systems. The big dust particles with the low movement velocity at the molecular regime of transportation, when the mean free path by the dust particles exceeds the characteristic dimensions of an air channel in the granular filtering medium in an air filter are most exposed to the self-structurization. This process results in the fast separation of small disperse coal dust particles from the air - dust aerosol flow process as experimentally observed in the sub-surface layer of granular filtering medium at the absorber's input at the initial time periods of *IAF's* operation. The structurization of small disperse coal dust particles is stipulated by the presence of big excess free energy in the small disperse coal dust phase of granulated material [3]. In the researched case, an interconnection between the dust particles can have place due to the direct nuclear or phase contacts [9], lowering the free energy of system. In the certain cases, the electrostatic forces can have an influence on the dust particles [4]. The direct contacts are characterized by the magnitude of force of up to $10^{-8}$ N, the phase contacts are characterized by the magnitude of force of $10^{-7}$ N per one contact [9]. The presence of these forces is enough to create the relatively strengthened structures. As it was explained above, the dust particles with the large dimensions are the first, which take place in the process of dust structurization, because they are exposed to the action by the small forces, and they have low velocities of movement during the air - dust aerosol flow. The dust structure, created by the dust particles with the large dimensions can be regarded as an effective air filter for the dust particles of similar dimensions. However, the free penetration of small dust fractions is possible through such an air filter. During the dust structurization process, there is a decrease of the characteristic scale of structure, resulting in an increase of the magnitude of centrifugal forces in an air channel in the granular filtering medium and the inclusion of the smaller dust fractions in the dust structurization process in view of the fulfillment of dust structurization criteria. The radius of an air channel $R$ decreases as the dust structure grows, leading to an increase of the aerodynamic resistance at the aerosol blow process in an air filter. The particular dependence $\Delta P(R)$ is defined by the relation between the laminar aerosol stream flow and the turbulent aerosol flow in an air channel. In the researched case, the turbulent aerosol flow in the granular filtering medium in an air filter is observed mainly.

### Modeling of air - dust aerosol flow and small disperse dust structurization in granular filtering medium with cylindrical coal absorbent granules in air filter.

The physical process of small disperse coal dust transportation is similar to the diffusion process with the complex dependence on the effective temperature. The effective temperature is the kinetic energy of dust particles divided by the *Boltzmann* constant $k_B$. The coefficient of diffusion has some physical meaning in this process. Let us formally write the diffusion equations for the dependence of dust concentration near the input boundary of absorber in an air filter and for the diffusion flow of air – dust aerosol stream $J$

$$\frac{\partial C}{\partial t} = D \frac{\partial^2 C}{\partial x^2}, \qquad (6)$$

$$J_i = -D_i \frac{\partial C_i}{\partial x} + C_i \upsilon_{F_i}, \qquad (7)$$

where $C$ is the total concentration of dust, which increases as time passes by; $C_i$ is the concentration of dust with the particles of the type of $I$; $x$ is the axis, directed inside an air filter; $D$ is the total coefficient of diffusion; $D_i$ is the diffusion coefficient of particles of the type of $I$; $\upsilon_{Fi}$ is the transportation velocity of dust particles of the type of $i$ at an action of the total force $F_i$.

The dust particles transportation velocity can be written as [10]

$$\upsilon_{F_i} = \frac{F_i \cdot D_i}{k_B \cdot T^*} \qquad (8)$$

where $F_i$ is the capturing force of dust particles of certain type, derived above. The solution of first equation in the close proximity to the sub-surface layer of granular filtering medium at the air filter's input can be written as in eq. (9)

$$C(z) = C_{Z=0}[1 - erfz] = C_{Z=0} erfz, \qquad (9)$$

where $erfz$ is the *Gauss* "integral of errors", $z = \dfrac{(3)^{1/2} \cdot (x-1)}{2(Dt)^{1/2}}$ in the considered case. The solution of the second equation can be written as the *Gauss* distribution of dust concentration in eq. (10)

$$C(x) \propto \{Q_0 / (\pi D_i t)^{1/2}\} \exp((x-x_0)^2/4D_i t), \qquad (10)$$

where $Q_0 = \int C(x,t)dx$. At the air filtering process, the maximum of *Gauss* distribution of dust masses concentration shifts along the length of granular filtering medium in an air filter, hence $x_0 \sim V_{Fi} t$.

In Fig. 3, the dependence of relative distribution of dust masses is shown [2]. It is created with the use of points, obtained at the measurements of total dust mass in the containers and sections of absorber at the completion of experiment with the application of the method of cubic splines, taking to the consideration the gradients of dust masses distribution at the transitions between the containers and sections. The total time of air blow is around $3,6 \cdot 10^4$ sec.

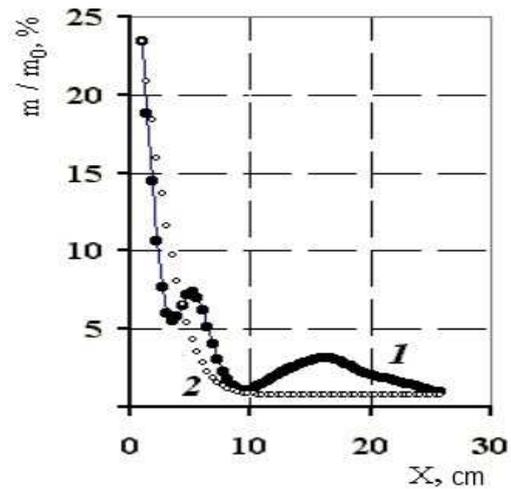

*Fig. 3. Distribution of relative coal dust mass as function of distance from absorber's input's surface in air filter(1 – experimental data, shown as percentage of total dust mass in absorber; 2 – computer modeling results, computed with application of eq. (7)).*

As it can be seen in Fig. 3, the small disperse coal dust particles distribution near the sub-surface layer at the input of absorber is well described by the diffusion approach with the certain common coefficient of diffusion $D$ (eq. 7). In this region, the small disperse dust particles with the maximal dimensions make a main contribution to the structurization process. As it was already explained, the dust particles with the large dimensions fulfill the empty spaces and create the initial dust precipitations structures, which experience the subsequent growth due to the participation of dust particles with the smaller dimensions during the aerosol blow in the air channels in the granular filtering medium, resulting in a significant decrease of the effective radius $R$ of air channels in an air filter. It is assumed that this physical process of small disperse coal dust precipitations accumulation is a main reason of aerodynamic resistance increase, which must appear at the laminar and turbulent limits.

As it follows from the data in Fig. 3, there are the two additional maximums of dust masses accumulation along the absorber's length, which can be connected with the nature of transportation of certain coal dust fractions with the small dimensions. In the case of $F_l<F_S$, the process of dust particles transportation is regulated by the complex relation between the relatively free movement with the possibility of aggregation on the surface of cylindrical coal granules and the subsequent capturing of dust particles by the aerosol flows. Such coal dust particles movement is analogous to the transfer of gases with the different molecular masses through the gas chromatographic column [11]. In this case, the light coal dust fractions have the high velocity of transportation, resulting in the fast movement through the chromatographic column. Let us note that the obtained experimental results confirm the described nature of dust transportation process [2].

The computer modeling of maximums of concentration of small disperse coal dust particles in the dependence of the relative coal dust mass accumulation on the distance from the absorber's input's surface was completed in the *Matlab*, using the *Gauss* distribution theory.

In Fig. 4, the experimental dependence, describing the distribution of relative coal dust mass as a function of the distance from the absorber's input's surface in the close proximity to the first peak's position in the granular filtering medium of absorber in an air filter, is presented. In Fig. 4, the experimental results curve is well approximated by the computer modeling curve, representing a solution of the diffusion equation with the following coefficients

$$C_1(x) = C_{01} + \frac{A_1}{w_1\sqrt{\pi/2}} \exp\left(-2\left(\frac{x-x_{01}}{w_1}\right)^2\right), \quad (11)$$

where $C_{01} = 1,17\ \%$, $A_1 = 24,12\ \%\ cm$, $x_{01} = 4,91\ cm$, $w_1 = 3,0725\ cm$.

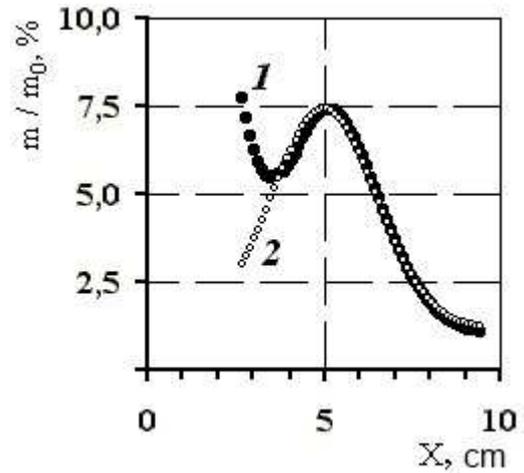

*Fig. 4. Distribution of relative coal dust mass as function of distance from absorber's input's surface in close proximity to first peak's position in granular filtering medium of absorber in air filter (1 - experimental data, 2 – computer modeling results).*

In Fig. 5, the experimental dependence, describing the distribution of relative coal dust mass as a function of the distance from the absorber's input's surface in close proximity to the second peak's position in the granular filtering medium of absorber in an air filter, is presented. In Fig. 5, the experimental results curve is also well approximated by the computer modeling curve, representing a solution of the diffusion equation with the following coefficients

$$C_2(x) = C_{02} + \frac{A_2}{w_2\sqrt{\pi/2}} \exp\left(-2\left(\frac{x-x_{02}}{w_2}\right)^2\right), \quad (12)$$

where $C_{02}=0,9579\ \%$, $A_2=18,422\ \%\ cm$, $x_{02}=16,341\ cm$, $w_2 = 6,8875\ cm$.

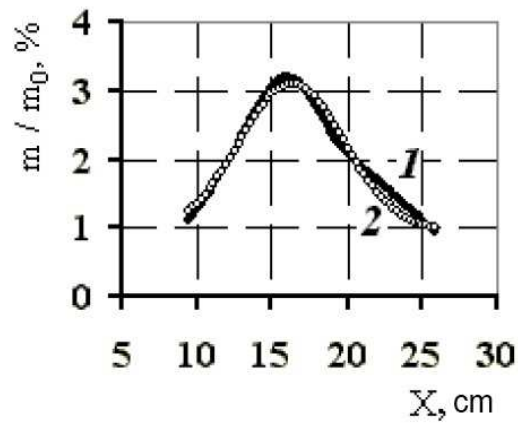

Fig. 5. *Distribution of relative coal dust mass as function of distance from absorber's input's surface in close proximity to second peak's position in granular filtering medium of absorber in air filter (1 - experimental data, 2 – computer modeling results).*

The researched distributions of relative coal dust masses have the different velocities of small disperse dust transportation and various effective temperatures, and they shift at an action by the capturing force during the air blow process in an air filter. Going from the presented formulas, it is possible to find an interconnection between the relative dimensions of dust particles, creating the first and second dust masses accumulation maximums. However it is necessary to make some additional suppositions about the character of velocities distribution and select the appropriate approaches to describe the nature of physical mechanisms of small disperse coal dust transportation in an air filter, with the purpose to make the meaningful mathematical estimations. Using the diffusion theory approach, the value of diffusion coefficient at the initial part of small disperse coal dust transportation process in an air filter is obtained: $D \approx 3.8 \cdot 10^{-5} \, cm^2/sec$. In the case of the distribution of relative coal dust masses as a function of the distance from the absorber's input's surface in the granular filtering medium in an air filter, the measured diffusion coefficients are: $D_1 \approx 3.29 \cdot 10^{-5} \, cm^2/sec$ for the 1st maximum, $D_2 \approx 1.647 \cdot 10^{-4} \, cm^2/sec$ for the 2nd maximum. As it can be seen, these diffusion coefficients have the values, which are significantly smaller (in the 3 - 4 orders of magnitude) than the diffusion coefficient of oxygen atoms in the air at the normal conditions, and correspond to the diffusion process by the air – dust aerosol in the granular filtering medium in an air filter.

## Conclusion.

As it is shown in this research, the transportation of small disperse dust precipitations in the granular filtering medium with the cylindrical coal absorbent granules in an air filter is a complex physical process, which is characterized by the following parameters: the dynamics of air – dust aerosol flow; the density of air – dust aerosol; and the nature of interaction between the small disperse dust particles and the geometry of elements of granular filtering medium in an air filter. The introduced physical criteria to characterize the physical mechanisms of high disperse dust masses transportation as a function of the dust particles dimensions, the dust particles transportation velocity, and the geometric dimensions of air channels in the porous granular filtering medium allow us to obtain the accurate physical characteristics of air filtering elements with the purpose of effective air cleaning. The new physical mechanisms, resulting in an appearance of the characteristic dependences of small disperse coal dust transportation along the adsorber in an air filter, are also proposed.

It makes sense to note that, in the case of the dynamic air blow, there can be the processes of coal dust structure destruction, when the magnitude of dynamic air trust forces exceeds the magnitude of dust structure hardness forces. This is one of the criterions, defining the operational stability of the full system. During the dynamic air blow process, the big fragments of coal dust structure are captured by the air-dust aerosol and transported along the granular filtering medium, experiencing the numerous collisions with the surfaces of the inner walls in the aerosol transportation channels. The dynamic air blow process is accompanied by an appearance of "the sand stream effect," when the magnitude of dust particles kinetic energy is bigger than the magnitude of work, which is necessary to overcome the adhesion forces. As a result, this physical process originates the sharp jettisoning of high disperse coal dust outside the air filter's absorbent containers. Thus, the cleaning of the filtering elements from the accumulated coal dust precipitations in the granular filtering medium in an air filter takes place. This type of the self-destruction of coal dust precipitations structure can limit the further decrease of dimensions of aerosol flow channels in the granular filtering medium during the air filter's operation in the regime of constant turbulent air flow. In this regime, the decrease of dimensions of effective cross-section of aerosol flow channels results in an appearance of a big gradient of air pressure with the subsequent increase of velocity of air – dust aerosol flow in the granular filtering medium in an air filter. Let us emphasis that the decrease of cross-sections of aerosol flow channels in the granular filtering medium in an air filter is limited by the system's stability criterion.

*E-mail: ledenyov@kipt.kharkov.ua

————————

1. L. I. Fedorova, P. Ya. Poltinin, L. V. Karnatsevich, M. A. Hazhmuradov, S. O. Lystsov, V. V. Teslenko, Yu. L. Kovrizhkin, Influence by contraction and mechanical wear of adsorber on aerodynamic parameters of coal adsorber of type of АУ-1500 in ventilation systems at nuclear power plants, *Nuclear Energy,* v. **82**, issue 4, pp. 279-283, 1999.
2. I. M. Neklyudov, L. I. Fedorova, P. Ya. Poltinin, L. V. Karnatsevich, Influence by features of accumulation of coal dust fraction in layer of adsorber on increase of aerodynamic resistance of coal iodine air filters in ventialation systems at nuclear power plant, *Problems of Atomic Science and Technology* (*VANT*), Series «*Physics of radiation damages and radiation materials*», №6 (84), pp. 65-70, 2003.
3. N. B. Ur'ev, Physical – chemical dynamics of disperse systems, *Uspekhi Khimii*, v. **79**, №1, pp. 39-62, 2004.
4. E. V. Fortov, A. G. Khrapak, V. I. Khrapak et al., Dust plasma, *Uspekhi Fizicheskih Nauk* (*UFN*), v. **174**, №5, pp. 495-544, 2004.
5. L. G. Loitsyansky, Mechanics of fluid and gas, *Nauka*, Moscow, 840 p., 1987.
6. L. D. Landau, E. M. Lifshits, Mechanics of continuous mediums, *GITTL*, Moscow, 788 p., 1953.

7. N. Z. Frenkel, Hydraulics, *Gosenergoizdat*, Moscow, 456 p., 1956.

8. A. M. Kolmogorov, On the logarithmically normal distribution of dimensions of particles at granulation, *DAN USSR*, v. **31**, №2, pp. 99-101, 1941.

9. P. A. Rebinder, Selected research works. Surface phenomena in disperse systems. Physical – chemical mechanics.*, Nauka*, Moscow, 244 p., 1979.

10. G. P. Stark, Diffusion in Solid States, *Energy,* Moscow, p. 240, 1980.

11. A. Kleimans, Chromatography of gases, *IIL*, Moscow, 320 p., 1959.

12. O. P. Ledenyov, I. M. Neklyudov, P. Ya. Poltinin, L. I. Fedorova, *Problems of Atomic Science and Technology* (*VANT*), Series «*Physics of radiation damages and radiation materials*», № 3(86), pp. 115-121, 2005, (in Russian); http://vant.kipt.kharkov.ua/ARTICLE/VANT_2005_3/article_2005_3_115.pdf